\documentclass[utf8]{frontiersFPHY} 

\usepackage{url,hyperref,lineno,microtype,subcaption}
\usepackage[onehalfspacing]{setspace}
\usepackage{dirtytalk}

\newcommand{\BIG}{{\sf BIG}}
\newcommand{\SLIM}{{\sf SLIM}}
\newcommand{\QUAINT}{{\sf QUAINT}}

\def\keyFont{\fontsize{8}{11}\helveticabold}
\def\firstAuthorLast{Vecchi {et~al.}} 
\def\Authors{M.~Vecchi\,$^{1,*}$, P.-I.~Batista\,$^{2}$, E.~F.~Bueno\,$^{1}$,  L.~Derome$^{3}$, Y.~G\'enolini$^{4}$, and D.~Maurin\,$^{3}$}
\def\Address{$^{1}$Kapteyn Astronomical Institute, University of Groningen, Landleven 12, 9747 AD Groningen, The Netherlands\\
$^{2}$Deutsches Elektronen-Synchrotron, Platanenallee 6, 15738 Zeuthen, Germany\\
$^{3}$LPSC, Universit\'e Grenoble Alpes, CNRS/IN2P3, 53 avenue des Martyrs, 38026 Grenoble, France\\
$^{4}$ Univ. Grenoble Alpes, USMB, CNRS, LAPTh, F-74940 Annecy, France}
\def\corrAuthor{M. Vecchi}
\def\corrEmail{m.vecchi@rug.nl}

\begin{document}
\onecolumn
\firstpage{1}

\title[Rigidity dependence of galactic cosmic-ray fluxes]{The rigidity dependence of galactic cosmic-ray fluxes and its connection with the diffusion coefficient} 

\author[\firstAuthorLast ]{\Authors} 
\address{} 
\correspondance{} 

\extraAuth{}

\maketitle
\begin{abstract}
Thanks to tremendous experimental efforts, galactic cosmic-ray fluxes are being measured up to the unprecedented percent precision level. The logarithmic slope of these fluxes is a crucial quantity that promises us information on the diffusion properties and the \textit{primary} or \textit{secondary} nature of the different species. However, these measured slopes are sometimes interpreted in the pure diffusive regime, guiding to misleading conclusions. In this paper, we have studied the propagation of galactic cosmic rays by computing the fluxes of species between H and Fe using the USINE code and considering all the relevant physical processes and an updated set of cross-section data. We show that the slope of the well-studied secondary-to-primary B/C ratio is distinctly different from the diffusion coefficient slope, by an offset of $\sim 0.2$ in the rigidity range in which the AMS-02 data reach their best precision (several tens of GV).
 Furthermore, we have demonstrated that none of the species from H to Fe follows the expectations of the pure-diffusive regime. We argue that these differences arise from propagation processes such as fragmentation, convection, and reacceleration, which cannot be neglected. On this basis, we also provide predictions for the spectral slope of elemental fluxes not yet analysed by the AMS collaboration. 


\section{}
\tiny
 \keyFont{ \section{Keywords:} astroparticle and fundamental physics,  cosmic rays-diffusion-methods, high-energy astrophysical phenomena, cosmic-ray nuclei, galactic cosmic particles, Phenomenology} 

\end{abstract}

\section{Introduction}
\label{sec:intro}
The Alpha Magnetic Spectrometer (AMS-02), taking data onboard the ISS for more than ten years, has been providing for the first time measurements of galactic cosmic-ray (CR) fluxes in the GV to TV range with percent level precision~\cite{2021PhR...894....1A}. These new results have revealed unexpected spectral features that challenge the theoretical framework used to describe the CR origin and propagation.

Cosmic rays can be divided into two broad classes: primary species, such as carbon and oxygen, are those accelerated at the sources, while secondary species, such as boron and lithium, are produced as a consequence of nuclear interactions of primary species during their propagation through the interstellar medium (ISM). The most promising observable to study the propagation of CRs in the Galaxy is studying secondary species or the flux ratio of a secondary species to a primary one. The measurement of the Boron-to-Carbon flux ratio, B/C in the following, has been widely used to test propagation scenarios. 
The first evidence for a break in the B/C was found by AMS-02~\cite{2016PhRvL.117w1102A,2018PhRvL.120b1101A}. This break most probably originates from a transition of diffusion regime rather than from source effects~\citep{2017PhRvL.119x1101G}. 

We develop on our preliminary study~\citep{Vecchi:2020nvb} and use the USINE code~\citep{2020CoPhC.24706942M} to show that the measured B/C slope does not directly represent the diffusion coefficient value in the AMS-02 range, and that the measured slopes for all species from H to Fe have non-trivial dependence with rigidity, especially for heavier elements.

\section{Method}
\label{sec:methods}
The transport in the Galaxy for a CR species of index $i$  can be described by means of the following equation, assuming steady-state~\cite{Amato:2017dbs}: 
\vskip 0.5 cm 
\begin{eqnarray}
\label{eq:prop}
& - \displaystyle\;\vec{\nabla}_{\bf x} \left\{ K(E)\,\vec{\nabla}_{\bf x}\Phi_i -
\vec{V}_{\rm c} \Phi_i \right \} 
+ \frac{\partial}{\partial E} \left\{ b_{\rm tot}(E)\;\Phi_i - \beta^2\, K_{pp}\,\frac{\partial \Phi_i}{\partial E} \right\}
+ \sigma_{i}\,v_i\, n_{\rm ism}\, \Phi_i + \Gamma_{i}\,\Phi_i \nonumber\\
& = \displaystyle\;q_{i}+\sum_{j} \left\{ \sigma_{j\to i}v_j n_{\rm ism}\,+\Gamma_{j\to i}
\right\}\,\Phi_j \;.
\end{eqnarray}
\vskip 0.5 cm 
This equation provides the spatial and energy evolution of the differential interstellar CR
density per unit energy $\Phi_i\equiv dn_i/dE$, assuming a net primary injection rate of
$q_i$, and an  injection rate for secondary species arising from inelastic processes converting heavier
species of index $j$ into $i$ species (with a production rate $ 
\sigma_{j\to i}v_j n_{\rm ism}$ on the ISM density $n_{\rm ism}$, or a decay rate
$\Gamma_{j \to i}$). The form of the spatial diffusion coefficient $K(E)$ will be described in detail in the following. The other processes are mainly relevant at low rigidity. However, they also affect the determination of the higher-energy parameters. The convection is described by a velocity $\vec{V}_{\rm c}$, the diffusive
reacceleration is parameterized by the energy-dependent coefficient $K_{pp}$, the particle velocity is indicated as $\beta$, and the inelastic
destruction rate is given by $\sigma_i v_i n_{\rm ism}$ with the  $\sigma$'s
being energy-dependent nuclear cross-sections; the rate characterizes energy losses
$b_{\rm tot}\equiv dE/dt$, which includes ionization and Coulomb processes  as well as adiabatic losses induced by convection, and a first-order term from reacceleration.

Following the work presented in~\cite{2019PhRvD..99l3028G}, our study is done in the frame of a 1D propagation model. In this model, CRs are confined in the magnetic halo described as an infinite slab in the radial direction and of half-height $L$. The value of $L$ is fixed to 5 kpc, and it was found to have a negligible impact on the results. The vertical coordinate $z$ is the only relevant spatial coordinate in this frame. The sources of CRs and the interstellar medium gas lie in the galactic disk, which has an effective half-height  $h = 100$~pc. Finally, the observer is located at z = 0. The diffusion coefficient is a crucial physical ingredient to describe CR transport as it represents the scattering of CRs off magnetic turbulence. We assume that it is a scalar function of rigidity, and that it is homogeneous and isotropic all over the magnetic slab.  
We follow the approach presented in~\cite{2019PhRvD..99l3028G}, where the diffusion coefficient includes a break in both the  low- and high-rigidity range:
\vskip 0.2 cm 
\begin{equation}
\label{eq:def_K}
K(R) = \underbrace{\beta^\eta}_{\text{non-relativistic~regime}} \, K_{10} \,
\underbrace{\left\{ 1 + \left( \frac{R}{R_{\rm l}} \right)^{\frac{\delta_{\rm l}-\delta}{s_{\rm l}}}
  \right\}^{s_{\rm l}}}_{\text{low-rigidity~regime}}\,
\underbrace{\left\{  \frac{R}{\left(R_{\rm 10}\equiv 10\,{\rm GV}\right)} \right\}^\delta}_{\text{intermediate~regime}} \,
\underbrace{\left\{  1 + \left( \frac{R}{R_{\rm h}} \right)^{\frac{\delta-\delta_{\rm h}}{s_{\rm h}}}
  \right\}^{-s_{\rm h}}}
_{\text{high-rigidity~regime}}\,.
\end{equation}
\vskip 0.2 cm 
We make use of the three propagation models, dubbed BIG, SLIM, and QUAINT, which were proposed in~\cite{2019PhRvD..99l3028G}. These models were found to provide an accurate description of the B/C data from AMS-02 and individual fluxes of Li, Be, B, and the helium isotopes~\cite{2020A&A...639A..74W}. 

For the nuclear production and spallation cross-sections, we use as reference the set of tables from the \emph{Galprop} package, following the approach described in~\citep{2019A&A...627A.158D} and recently updated as in~\citep{Maurin:2022irz}. We generate the fluxes of CR nuclei using the source spectrum parameters discussed in~\citep{2020PhRvR...2b3022B} and the propagation parameters described in ~\citep{2019PhRvD..99l3028G} and~\citep{2020A&A...639A.131W}. The calculated fluxes are Top-of-Atmosphere quantities modulated with the force field approximation. We obtain the modulation potential $\phi=670$~MV from the neutron monitor data using the dedicated feature, available on the Cosmic Ray Data Base~\cite{2014A&A...569A..32M,Maurin:2020teo}, based on~\cite{Ghelfi:2016pcv}.

The slope of the diffusion coefficient is given by:
\begin{equation}
    S_{K} = \text{d}\left[\log{K(R)}\right]/\text{d}\left[\log{R}\right] = - S_{1/K},
\label{formula:k}
\end{equation}
The slope of the measured CR flux  $S_{\Phi}$ is given by: 
\begin{equation}
    S_{\Phi} = \text{d}\left[\log{\phi}\right]/\text{d}\left[\log{R}\right],
\label{formula:gamma}
\end{equation}
We use the following quantity to investigate the difference between the slope of the measured B/C and the slope of the diffusion coefficient:
\begin{equation}
    \Delta_S = S_{B/C}  - S_{1/K},
\end{equation}

As mentioned in the introduction, we also want to investigate the slope of the primary and secondary species. To do so, we define, for all CR species from hydrogen to iron, the following slope ratio:
\begin{equation}
    \Delta_{\gamma} = \frac{S_{\Phi}-(\alpha(Z)+S_{1/K})}{S_{1/K}}\;.
    \label{eq:Deltagamma}
\end{equation}
where $\alpha(Z)$ describes  the source spectral indices. In the purely diffusive regime, we 
recall that the slopes of pure primary species are $S_{\Phi} = S^{\rm prim}=-\alpha(Z)-S_K(R)$ implying $\Delta_{\gamma}=0$, while for pure secondaries $S_{\Phi} = S^{\rm sec}=-\alpha(Z)-2S_K(R)$, implying $\Delta_{\gamma}=-1$.
In reality, all primary species have some fraction of secondaries. To quantify it, it is useful to introduce the fraction $f_{\rm prim}$ of primary origin in the total flux, given by: $f_{\rm prim}=\frac{\Phi_{\rm prim}}{\Phi_{\rm sec} + \Phi_{\rm prim}}$, with the elemental flux split as $\Phi=\Phi_{\rm prim}+\Phi_{\rm sec}$, into a pure primary and a pure secondary component. In this work we assumed Li, Be and B to be pure secondary species.
\section{Results}
\label{sec:results}
We discuss in this section our results, based on fluxes (or ratios) calculated as described above. For all our calculated slopes, we choose not to show the associated uncertainties (deriving from the transport parameter uncertainties and correlations) and mainly focus on the \SLIM{} propagation configuration. Indeed, these uncertainties are expected to be at the percent level, but they could reach up to tens of percent at TV rigidities. They neither impact the features we wish to put forth nor the conclusions we reach here, and they would only complicate the readability of the figures.

\subsection{B/C vs diffusion slope: has the diffusive regime been reached in AMS-02 data?}
\label{subsec:B/C}
At high rigidity enough, secondary-to-primary ratios are inversely proportional to the diffusion coefficient $K(R)$. Indeed, this rigidity-dependence is exact when diffusion becomes the dominant process in the propagation of cosmic-rays: inelastic interactions, energy losses, convection, and reacceleration processes compete with diffusion at low and intermediate energies (see, e.g. App.~D in \citealt{2019A&A...627A.158D}) and break down this exact scaling. Solar modulation also impacts the slope up to hundreds of GV.

The diffusive regime is particularly appealing since it allows to directly read the diffusion slope from the B/C data, without any modelling. The diffusion coefficient, in turn, gives us indications on the underlying magnetic turbulence from which diffusion arises \citep{2020Ap&SS.365..135M}. 
\begin{figure}[t!]
\begin{center}
\includegraphics[width=0.8\columnwidth]{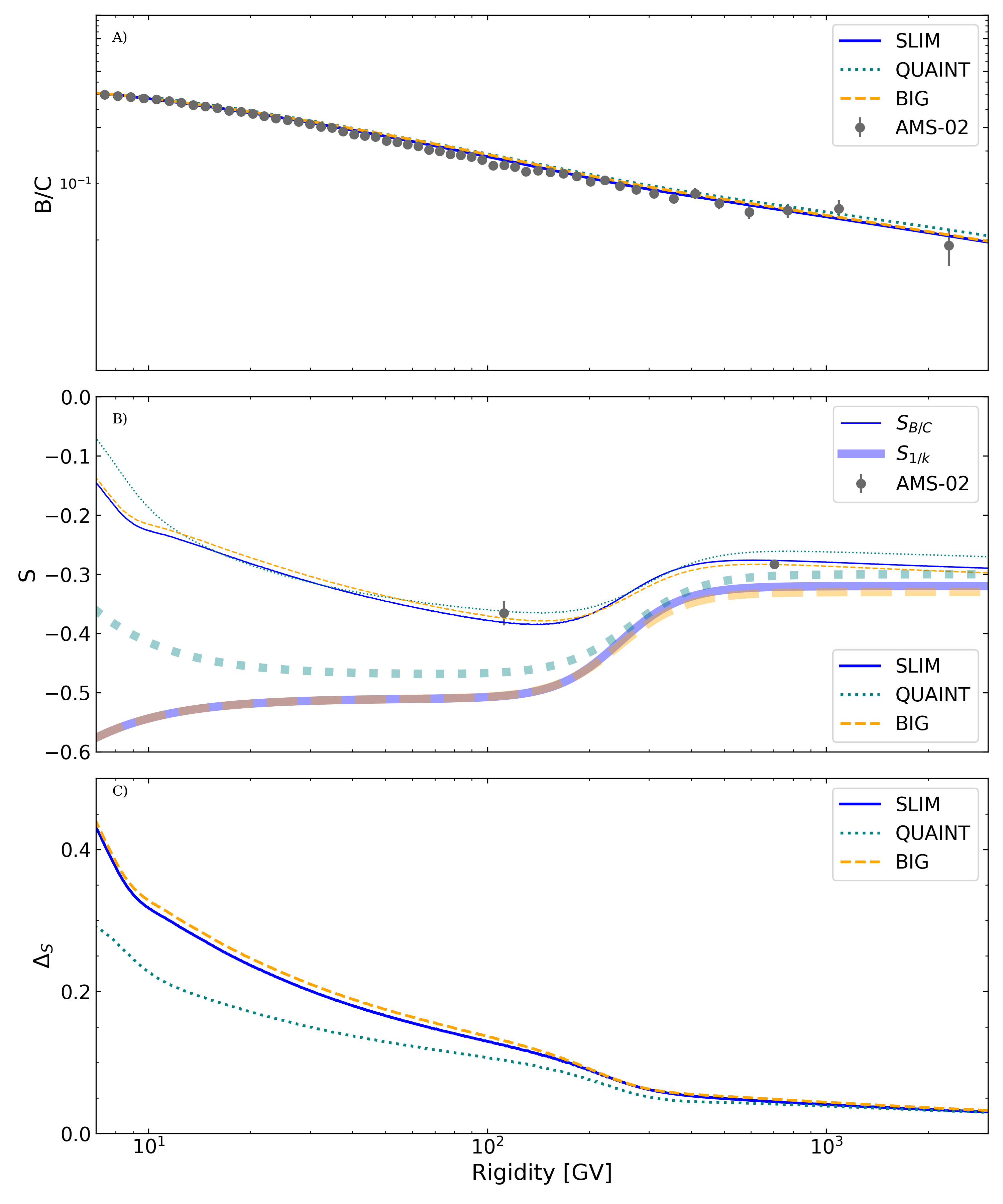}
\end{center}
\caption{Shown as a function of rigidity and for the three configurations \SLIM{} (blue solid lines), \BIG{} (orange dashed lines), and \QUAINT{} (cyan dotted lines) are (from top to bottom): \textbf{(A)} our B/C calculations (lines) and AMS-02 data \citep{2018PhRvL.120b1101A}; \textbf{(B)} logarithmic slopes $S_{\rm B/C}$ for the above B/C calculations (thin lines) and for AMS-02 data at two rigidities \citep{2018PhRvL.120b1101A}, along with the logarithmic slope $S_{1/K}$ for the inverse of the associated diffusion coefficient (thick lines); \textbf{(C)} $\Delta_s=S_{\rm B/C}-S_{1/K}$, the difference between the B/C and $1/K(R)$ logarithmic slopes.}
\label{fig:BC}
\end{figure}
One common misconception in literature is about when the purely diffusive regime is reached. Figure~\ref{fig:BC} shows, for the propagation configurations \BIG{}, \SLIM{}, and \QUAINT{} discussed in the previous section (Sect.~\ref{sec:methods}), three different panels highlighting the salient differences between the B/C and $1/K(R)$ slopes, as a function of rigidity.
The top panel \textbf{(A)} illustrates that all these three configurations fit well the AMS-02 B/C data \citep{2019PhRvD..99l3028G,2020A&A...639A.131W}.
The middle panel \textbf{(B)} shows the corresponding slopes of these B/C calculations (thin lines)---the slopes in all model configurations are similar since all calculations need to match the same B/C data. 
The modelled B/C slopes steadily decrease up to the high-rigidity break around 300~GV; beyond, a constant slope is slowly reached (above hundreds of TV). We immediately see that our best-fit diffusion coefficient slopes (thick lines) never matches the B/C slopes, except at the highest rigidities, although the same overall behaviour is observed (low- and high-rigidity change of slope). We stress that the behaviour is slightly different in \QUAINT{} (cyan thick dotted line) owing to the presence of reacceleration (absent in \SLIM{} and \BIG{} best-fit configurations). At any rate, the diffusion coefficient at intermediate and high-rigidity regimes are constant  
with $\delta_{10-100~\rm GV}=0.5 \pm 0.02$  \citep{2020A&A...639A.131W} and $\delta_{>300~\rm GV}=0.3\pm0.15$ \citep{2019PhRvD..99l3028G}. These results favour a Kraichnan turbulence regime~\cite{1965PhFl....8.1385K} at intermediate rigidity and are also compatible with a Kolmogorov turbulence~\cite{1941DoSSR..30..301K} spectrum after the rigidity break. These values are also consistent with the fact that the B/C slope for the AMS-02 data, shown in filled circles (in agreement with our B/C slope calculations in thin lines), is $S_{B/C\,\in\,[60.3-192]\rm~GV}\sim 0.36\pm0.02$ and $S_{B/C\,\in\,[192-3300]\rm~GV}\sim 0.28\pm0.04$ (see Fig.~3 of \cite{2018PhRvL.120b1101A}). In passing, we stress that the difference of B/C slopes between these two rigidities does not happen in the purely diffusive regime. This difference cannot be directly translated as the difference between the slopes of the underlying magnetic turbulence regimes.
Finally, the bottom panel \textbf{(C)} shows the difference between the B/C data and $1/K(R)$ slopes, i.e. the difference between the two families of curves (thick and thin) shown in the middle panel. In the intermediate rigidity range, we see the slow convergence to the purely diffusive regime ($\Delta_S=0$). We directly quantify how to recover the diffusion slope from the B/C data slope from these curves. Indeed, with $S_{K(R)} = |S_{\rm B/C}|+\Delta_S$, we see that we need to add to the slope measured in the data the value $\Delta_S\approx 0.15$ at $\sim 100$~GV and still $\Delta_S\approx 0.05$  at a few TV.
Our results show that a propagation model is needed to extract the spectral index of the magnetic turbulence from the slope of secondary-to-primary ratios. Indeed, the diffusive propagation regime is asymptotically reached at several hundreds of TV only. Assuming the B/C slope directly provides the slope of the diffusion coefficient strongly biases the conclusions drawn on the turbulence type; this bias grows with decreasing rigidities (at which the B/C slope is evaluated).

\subsection{Understanding the behaviour of logarithmic slopes ($Z=1-26$) vs rigidity}
\label{subsec:deltagamma}

Considering now the fluxes, it is interesting to see which propagation processes shape them. Their logarithmic slope is usually a good indicator of their primary (acceleration from material at source only) or secondary (produced via nuclear interactions of primary species during the transport only) origin. However, as for the B/C ratio, the diffusive regime is only asymptotically reached for very high rigidities. Furthermore, inelastic interactions, which play an important part at intermediate rigidities, roughly scale as $A^{2/3}$ (where $A$ is the atomic number), so that a growing impact on the fluxes (hence the slopes) is expected for growing atomic numbers.

\begin{figure}[t!]
\begin{center}
\includegraphics[width=0.9\columnwidth]{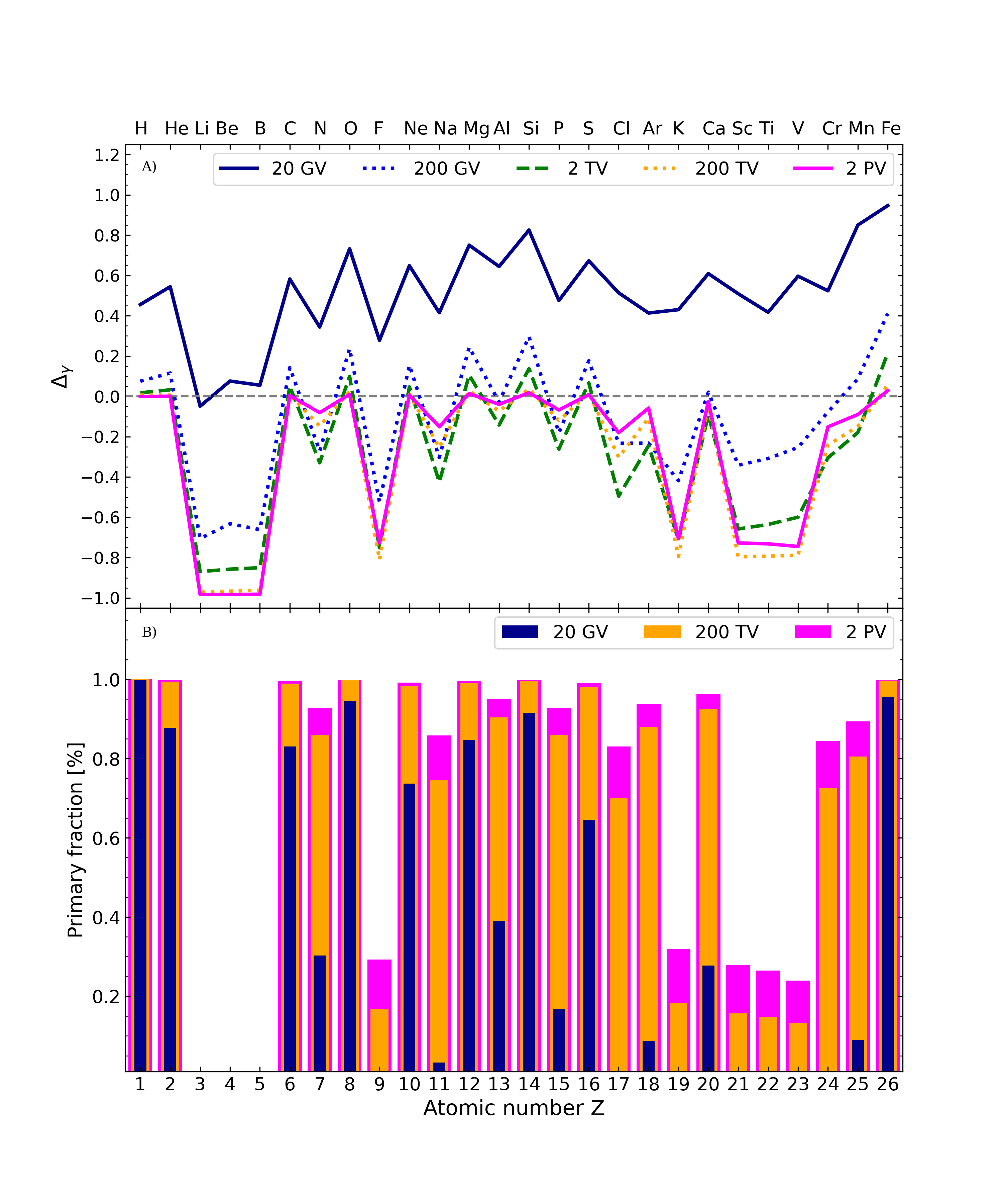}
\end{center}
\caption{Shown for the \SLIM{} model as a function of the charge $Z=1-26$ (from H to Fe) and for various rigidities (colour-coded with different line styles) are: \textbf{(A)} $\Delta_\gamma$ (see Eq.~\ref{eq:Deltagamma}), the  `normalised' difference between the flux and the `source + diffusion' logarithmic slopes; \textbf{(B)} primary fraction (relative to the total flux) for the elements.}
\label{fig:Deltagamma}
\end{figure}

The logarithmic slopes, or rather the quantity $\Delta_\gamma$ (see Eq.~\ref{eq:Deltagamma}) is displayed in the top panel \textbf{(A)} of Fig.~\ref{fig:Deltagamma} as a function of $Z=1-26$ (i.e., for H up to Fe) and for various rigidities (shown as different line styles and colours). As discussed in Sec.~\ref{sec:methods}, the definition of $\Delta_\gamma$ allows to factor out the presence of the diffusion break (similar for all species) and the different source spectral indices $\alpha(Z)$~\citep[e.g.,][]{2020PhRvR...2b3022B}. 
In the purely diffusive regime, we recall that the slopes of primary and  secondary species are $S^{\rm prim}=-\alpha(Z)-S_K(R)$ and $S^{\rm sec}=-\alpha(Z)-2S_K(R)$, which translates into $\Delta_\gamma^{\rm prim}=0$ and $\Delta_\gamma^{\rm sec}=-1$. If we first look at the slopes at 20~GV (solid dark blue line), i.e. in a regime where other propagation processes are significant, we have $\Delta_\gamma>0$ for all species. This means that the slopes of the measured fluxes, $S_{\rm data}$, are always softer than those expected in the purely diffusive regime, i.e. $|S_{\rm data}^{\rm prim,\,sec}|<|S_{\rm pure~diff.}^{\rm prim,\,sec}|$. This behaviour is expected since energy losses, inelastic interactions and Solar modulation effects (but also convection and reacceleration if present) are known to transfer particles from high to low rigidity in the two first decades measured by AMS-02. However, on 
the 20~GV curve, we can only unambiguously identify Li, Be, and B as secondary species, owing to the $\sim0.5$ drop in their $\Delta_\gamma$ values (compared to the neighbour species He and C). 
Indeed, for growing $Z$, the impact of inelastic cross-sections is also rising, translating into growing values of $\Delta_\gamma$ and causing an increasing difficulty to identify which species are of primary or secondary origin.

The quantity $\Delta_\gamma$ reaches the expected purely diffusive values at higher rigidities: at 200~GV, we already see primary species closing on $\Delta_\gamma=0$ and secondary species on $\Delta_\gamma=-1$ (except for $Z\gtrsim20$ where inelastic interactions still have a significant impact). In principle, for the highest rigidity shown, i.e. 2~PV (where the asymptotic diffusive regime holds), the heavy species should also converge to 0 or -1. However, there is no such thing as a pure primary or pure secondary species. We recall that secondary species have softer spectra (extra diffusion slope) than primary species. As a result, the fraction $f_{\rm prim}$ of primary origin in the total flux is a growing function of rigidity (except for Li, Be, and B whose primary fraction was explicitly set to zero in the calculation). This is illustrated in the bottom panel \textbf{(B)} of Fig.~\ref{fig:Deltagamma} for three rigidities.
The rigidity dependence of $\Delta_\gamma$ thus results from the competition of inelastic interactions and primary content of the species. For instance, most species have their values shifted downward for growing rigidities. However, species whose primary fraction significantly changes above 20~GeV, e.g. the mixed-species N ($Z=7$) with $f_{\rm prim}$ going from $\sim30\%$ at 20~GV to $\sim80\%$ at 200~TV, have their values shifted upwards above 2~TV. For the largest $Z$, where the impact of inelastic interactions is the strongest and where most $f_{\rm prim}$ also  significantly vary, the evolution of $\Delta_\gamma$ becomes non-trivial: this is illustrated, for 
instance, by the behaviour of species with
$Z=21-23$, called sub-Fe, whose $\Delta_\gamma$ values first decrease with rigidity (decreasing impact of destruction), but then increase above 200~TV (growing fraction $f_{\rm prim}\gtrsim20\%)$ of primaries in the flux). We stress that two opposite effects, namely the rigidity-dependent primary fraction and the rigidity- and $Z$-dependent impact of destruction on the fluxes, complicate the interpretation of the slopes of the nuclear species, except for almost `pure' primary species (H, O, Si, Fe) and `pure' secondary light species (Li, Be, B, and partly F). These effects should lead to non-trivial patterns for the slope of $Z>14$ elements (see next section); these elements should be analysed soon by the AMS collaboration. A last impacting effect we did not discuss here is the existence of contributions from multi-step fragmentation \citep{2018PhRvC..98c4611G}, i.e. the fact that a species $N_1$ breaks up into $N_2$, which itself can break up into $N_3$. This leads to tertiary contributions (extra slope $S_K(R)$) that could further impact the slope of some of these heavy species up to several tens of GV.

\subsection{Comparison to AMS-02 data for $Z\leq14$ and expectations for $Z=15-25$ elements}
\label{subsec:slopes_vs_AMS}

Now that we better understand the factors that drive the rigidity dependence of $\Delta_\gamma$ (see previous subsection), we can return to the rigidity dependence of the measured flux slope, $S_\Phi(R)$. We stress that the presence of the $\sim 300\;$GV high-rigidity break is no longer factored out, but that the salient features seen in $\Delta_\gamma$ should remain, that is: (i) the slope of most elements should be a fast decreasing function of $R$ below the break, but a slowly decreasing one (towards the pure diffusive regime value) above the break; (ii) asymptotically, `pure' primary and secondary species should be shifted by $S_K(R\to\infty)\approx0.3\pm0.15$ \citep{2019PhRvD..99l3028G}, see discussion in Sect.~\ref{subsec:B/C}; (iii) the growing impact of inelastic interactions with $A$ should be visible, especially at low rigidities, as decreasing slopes with $Z$; (iv) the impact of a growing primary fraction should be seen as a slope starting close to the slopes of `pure' secondary species (e.g., Li, Be, B) and moving to those of `pure' primary species (e.g., O, Si, Fe).

\begin{figure}[t!]
\begin{center}
\includegraphics[width=\columnwidth]{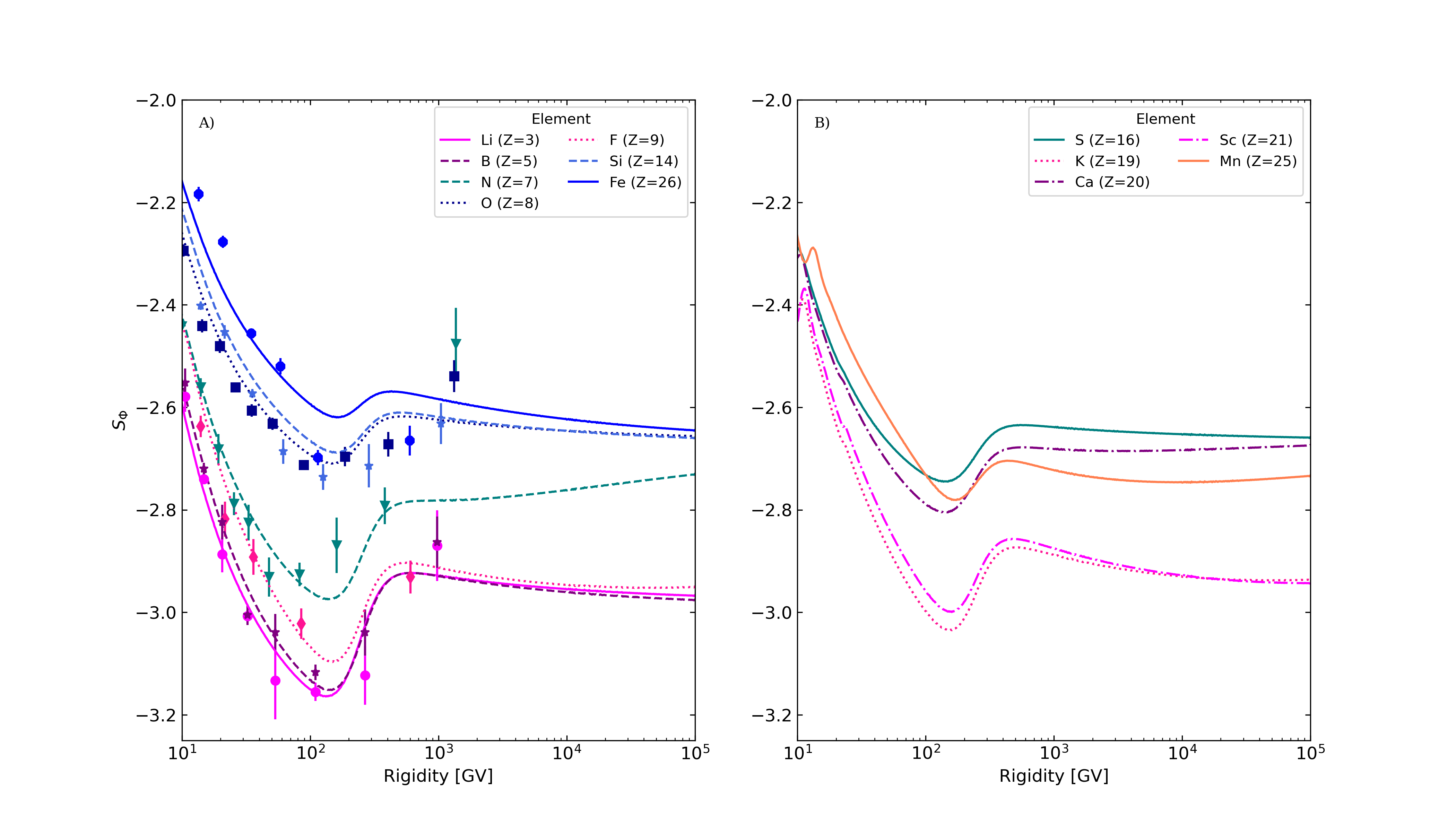}
\end{center}
\caption{The two panels show logarithmic slopes of TOA fluxes: in the left panel \textbf{(A)}, for a selection of elements already analysed by the AMS-02 collaboration (symbols, \citealt{2021PhR...894....1A}); in the right panel \textbf{(B)}, for a selection of elements accessible to the AMS-02 detector. The various lines correspond to our calculation in the \SLIM{} configuration. See text for discussion.}
\label{fig:slopesvsR}
\end{figure}
We show the flux slopes $S_\Phi(R)$ obtained in our calculations in  Fig.~\ref{fig:slopesvsR}. A similar slope dependence is shown for instance in \cite{2021PhRvD.103l3010S}. They are compared to the slopes of a selection of elements published by the AMS-02 collaboration \citep{2021PhR...894....1A} in the left panel \textbf{(A)}. We also show our predictions for a selection of elements not yet analysed (i.e. among $Z=15-25$) in the right panel \textbf{(B)}.

Focusing first on the left panel \textbf{(A)}, we find that the modelled slopes follow the behaviour we just recalled:~(i) Li ($Z=3$), B ($Z=5$) and F ($Z=9$), which have $f_{\rm prim}<5\%$ up to tens of TV, are ordered and shifted (i.e. decreasing  $S_\Phi(R)$) according to their growing destruction cross-section, before all slopes converge to the `universal' secondary flux slope;
(ii) the same ordering is observed for O ($Z=8$), Si ($Z=14$), and Fe ($Z=26$), though these species now converge towards the `universal' primary flux slope in our model; (iii) the pattern of a mixed species, N ($Z=7$) is striking, as its flux slope starts close to the `pure secondary' group and ends up close to the `pure primary' group. We stress that if N were to have a negligible primary fraction at tens of GV, its slope would be between B and F. Overall these predictions are in very good agreement with the AMS-02 data (symbols), though our model is clearly much below for the last N rigidity point, while the shape for Fe does not match the data above 100~GV (we briefly return to these issues below). In our model, as reflected by the data, the source slope for H and He are taken to be different from that of all other nuclei \cite{2020PhRvR...2b3022B}. As a result, their flux slope would be shifted (compared to other primary elements), and for readability reasons, we chose not to show them in the figure.

Focusing on the right panel \textbf{(B)}, that is, species in $Z=15-25$ not analysed by AMS-02 yet, we see a similar trend between `mostly' primary and `mostly' secondary species. Indeed, looking back at the bottom panel of Fig.~\ref{fig:Deltagamma}, we see that the elements K ($Z=19$) and Sc ($Z=21$) have a primary fraction very similar to that of F (i.e negligible at low and intermediate rigidities). The same should be the case of Ti and V (not shown for readability), belonging to the so-called sub-Fe group ($Z=21-23$).
The slope of these elements thus follows a similar pattern as F, though with a smaller slope at low rigidity, owing to the larger impact of inelastic interactions on this heavier element. On the other hand, Ca ($Z=20$) already has a primary fraction of 30\% at 20 GV, i.e. similar to that of N. The last two other elements shown, S ($Z=16$) and Mn ($Z=25$), fall between K (and Sc) and Ca in terms of their primary content---they have a $\sim10\%$ primary fraction at 20~GV (see bottom panel of Fig.~\ref{fig:Deltagamma})---, so that they are elements in which the effect of destruction and the impact of their primary content are maximally mixed to shape the slopes. Beside the fact that these flux slopes will converge to the asymptotic value of a primary flux, the fine details are very sensitive to the exact tiny and uncertain primary content of these elements (AMS-02 data should help constrain or set upper limits on these source terms).

\begin{figure}[t!]
\begin{center}
\includegraphics[width=0.8\columnwidth]{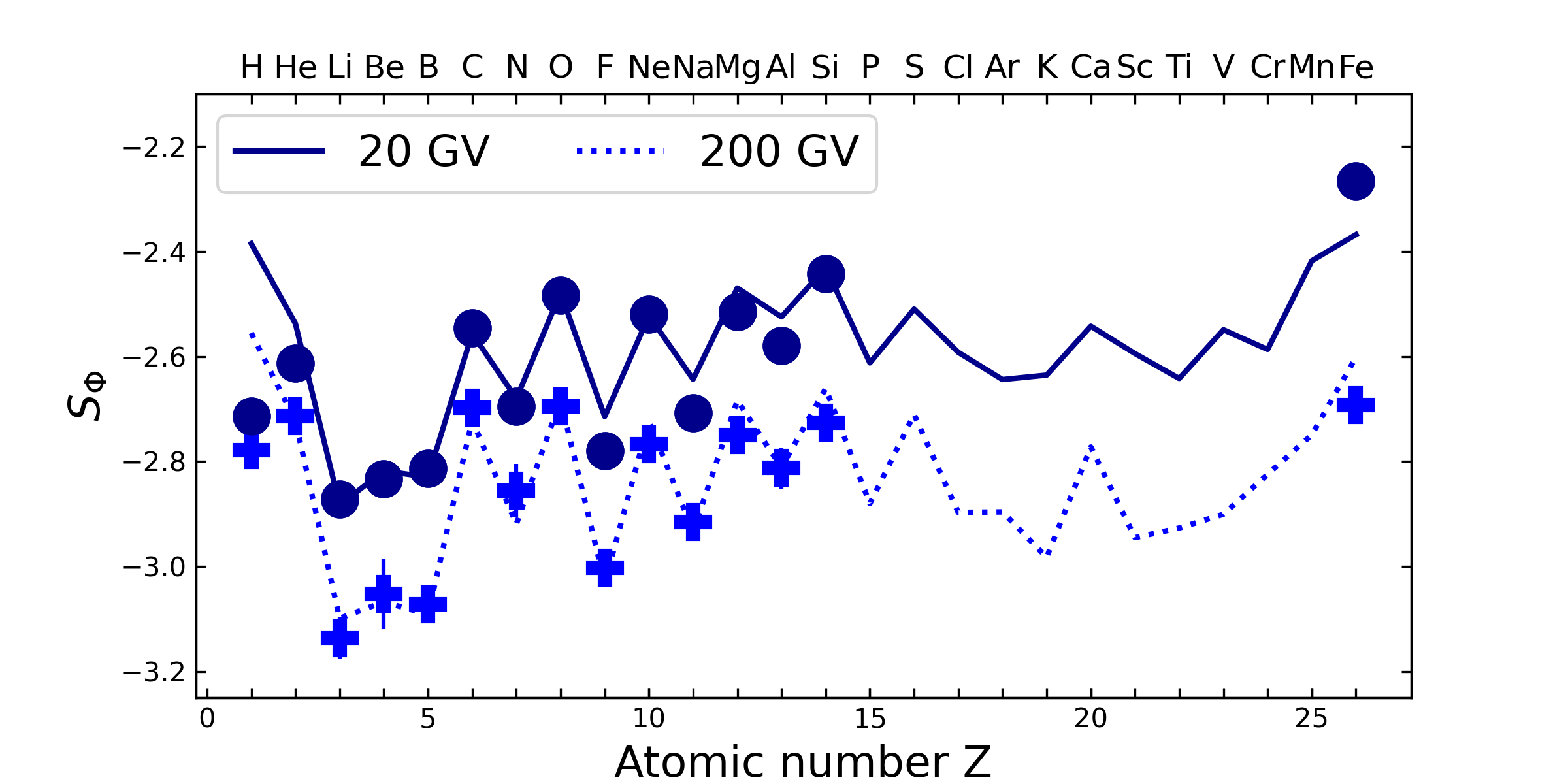}
\end{center}
\caption{Logarithmic slope at 20~GV (solid dark blue line) and 200~GV (dotted blue line) of our TOA flux calculation for $Z=1-26$ elements. Slopes calculated from AMS-02 data points \citep{2021PhR...894....1A} are shown as symbols at the corresponding energies.}
\label{fig:slopesvsZ}
\end{figure}

A final view of the trends and patterns in the flux slopes are illustrated by its $Z$ dependence shown in Fig.~\ref{fig:slopesvsZ}. This plot resembles the top panel of Fig.~\ref{fig:Deltagamma}, but instead of $\Delta_\gamma(Z)$, we now show $S_\Phi(Z)$ at only two rigidities, namely 20~GV (solid line) and 200~GV (dotted line): we do not show higher rigidities (for the sake of readability), as the associated slope values would lie between these two curves (because of the diffusion break at $\sim 300$~GV, see in Fig.~\ref{fig:slopesvsR}). As in the previous plot, we find a very good agreement between the modelled slopes and the measured ones (except for H and Fe); for the value of the experimental slopes, we use a simple interpolation for the desired rigidities from the published AMS-02 slopes \cite{2021PhR...894....1A}.

We recall that our models \citep{2019PhRvD..99l3028G,2020A&A...639A.131W} do not explicitly fit the source spectrum and source abundances. Rather, they assume a rigidity power-law dependence for all species, three different source spectral indices (one for H, one for He, and one for all other nuclei), and merely rescales elemental abundances on existing data at a single energy point. For this reason, it is not surprising, given the minimal number of ingredients used in the model, that some differences exist for some species. In particular, our models fail to match well low-rigidity protons and iron (see also \citep{2021PhRvD.103l3010S,2021ApJ...913....5B} who face similar difficulties for Fe). The origin of the high-rigidity break (and above) is still under scrutiny, and also the regions where AMS-02 data uncertainties are the largest. Whether the discrepancy between our models and the data at these high rigidities is due to a limitation of our models or statistical fluctuations in AMS-02 data are still to be investigated.
For example, the modelled slopes have significant uncertainties at high rigidities because the diffusion break parameters are not well constrained (because of larger uncertainties on the data).

\section{Discussion}
\label{sec:conclusions}

The best route to interpret the exquisite data collected by the AMS-02 experiment is to fit fluxes and ratios with propagation models. However, because (i) the source spectra and the diffusion coefficients are expected to be (or close to be) power-laws in rigidity, and (ii) in the purely diffusive regime, these fluxes and ratios are simple combinations of the two above power-laws, it is tempting to bypass the use of a propagation model and directly deduce their power-law index from the logarithmic slope of the measured fluxes and ratios. Moreover, it is also tempting to conclude on the primary or secondary origin of CR elements from the rigidity dependence of their flux slope.

We showed and stress that the purely diffusive propagation regime is only reached above hundreds of TV. The first consequence is that AMS-02 data, going up to a few TV at most, cannot be used to conclude on the source slope or diffusion slope without an underlying propagation model. For instance, a B/C slope of $1/3$ at a few hundred GV does not mean that the diffusion coefficient and the underlying magnetic turbulence are Kolmogorov-like. Our analysis indicates that at these rigidities, other propagation processes lead to a slope $\sim 0.2$ away from that of the pure diffusive asymptotic regime, so that AMS-02 data actually favour a Kraichnan-like turbulence (at intermediate rigidities). These effects (inelastic interactions, convection, reacceleration, and also Solar modulation) still impact the slope of secondary-to-primary ratios and fluxes around the rigidity break ($\sim 300$~GV). For this reason, it is also difficult to directly link the slope (before and after the break) to the two fundamental quantities that are the source and diffusion slope.

We also discussed in this paper the rigidity dependence (and $Z$ dependence) of the flux slopes. 
The general trends for all elements are the following:~(i) decreasing slopes due to the transition from `all propagation effect matter' to the diffusion-dominated regime, (ii) shallower slopes for heavier nuclei because of their larger inelastic cross-sections, (iii) overall shift between primary and secondary species. However, we highlighted that the competition between inelastic interactions (growing with $Z$) and primary content of the elements (growing with $R$) leads to non-trivial dependencies of the flux slopes, especially for the awaited AMS-02 measurements for $Z=15-25$. Indeed, two families of slopes should be observed: the `almost purely' secondary species (K, Sc, Ti, and V) with a shallower slope, and the remaining elements (whose primary content varies between a few percent up to 30\% at 20 GV) with steeper and more dispersed slopes.

Measured flux slopes have non-trivial behaviours, but they remain interesting to show for the following reasons: (i) qualitative information on rigidity breaks can be obtained without any underlying propagation model (breaks are more easily seen in the logarithmic slopes than in the flux themselves); 
(ii) as underlined above, whether CR elements are (almost purely of) primary or secondary origin can be roughly concluded from their observed slope difference $(\gtrsim 0.3$ above $\sim 100$~GV); 
(iii) finally, whether the measured slopes reach the expected asymptotic regime could be a useful validation of the model or of its limitation. For instance, stochasticity of the sources impacts the rigidity dependence of the fluxes \cite{2017A&A...600A..68G}, though with a negligible impact below several tens or TV \cite{2017EPJWC.13602006S,2019ApJ...887..250P,2021PhRvD.104l3029E}.
At these energies (and higher), interesting information should be brought by forthcoming data of space calorimeters such as DAMPE \cite{DAMPE:2017cev}, CALET \cite{Torii:2019wxn} and NUCLEON \cite{Atkin:2017vhi}.


\section*{Conflict of Interest Statement}

The authors declare that the research was conducted in the absence of any commercial or financial relationships that could be construed as a potential conflict of interest.

\section*{Author Contributions}
MV took the lead on the research work by running the simulations, performing most  of the analysis and producing all the figures. PB, EB, LD and DM contributed to the code development. MV, DM and YG wrote substantial parts of the manuscript. All authors contributed equally to the discussions, read the manuscript and provided critical feedback. 

\section*{Funding}
MV acknowledges the financial support of the University of Groningen.

\section*{Acknowledgments}
MV acknowledges the valuable discussion and encouragement from Mathieu Boudaud during the early phases of this work.  
\bibliographystyle{frontiersinHLTH&FPHY} 
\bibliography{frontiers}

\begin{thebibliography}{29}
\expandafter\ifx\csname natexlab\endcsname\relax\def\natexlab#1{#1}\fi
\expandafter\ifx\csname urlstyle\endcsname\relax
  \expandafter\ifx\csname doi\endcsname\relax
  \def\doi#1{doi:\discretionary{}{}{}#1}\fi \else
  \expandafter\ifx\csname doi\endcsname\relax
  \def\doi{doi:\discretionary{}{}{}\begingroup \urlstyle{rm}\Url}\fi \fi
\expandafter\ifx\csname selectlanguage\endcsname\relax
  \def\selectlanguage#1{}\fi

\bibitem[{{Aguilar} et~al.(2021){Aguilar}, {Ali Cavasonza}, {Ambrosi},
  {Arruda}, {Attig}, {Barao} et~al.}]{2021PhR...894....1A}
{Aguilar} M, {Ali Cavasonza} L, {Ambrosi} G, {Arruda} L, {Attig} N, {Barao} F,
  et~al.
\newblock {The Alpha Magnetic Spectrometer (AMS) on the international space
  station: Part II - Results from the first seven years}.
\newblock {\em \physrep\/} {\bf 894} (2021) 1--116.
\newblock \doi{10.1016/j.physrep.2020.09.003}.

\bibitem[{{Aguilar} et~al.(2016){Aguilar}, {Ali Cavasonza}, {Ambrosi},
  {Arruda}, {Attig}, {Aupetit} et~al.}]{2016PhRvL.117w1102A}
{Aguilar} M, {Ali Cavasonza} L, {Ambrosi} G, {Arruda} L, {Attig} N, {Aupetit}
  S, et~al.
\newblock {Precision Measurement of the Boron to Carbon Flux Ratio in Cosmic
  Rays from 1.9 GV to 2.6 TV with the Alpha Magnetic Spectrometer on the
  International Space Station}.
\newblock {\em Physical Review Letters\/} {\bf 117} (2016) 231102.
\newblock \doi{10.1103/PhysRevLett.117.231102}.

\bibitem[{{Aguilar} et~al.(2018a){Aguilar}, {Ali Cavasonza}, {Ambrosi},
  {Arruda}, {Attig}, {Aupetit} et~al.}]{2018PhRvL.120b1101A}
{Aguilar} M, {Ali Cavasonza} L, {Ambrosi} G, {Arruda} L, {Attig} N, {Aupetit}
  S, et~al.
\newblock {Observation of New Properties of Secondary Cosmic Rays Lithium,
  Beryllium, and Boron by the Alpha Magnetic Spectrometer on the International
  Space Station}.
\newblock {\em \prl\/} {\bf 120} (2018a) 021101.
\newblock \doi{10.1103/PhysRevLett.120.021101}.

\bibitem[{{Génolini} et~al.(2017{\natexlab{a}}){Génolini}, {Serpico},
  {Boudaud}, {Caroff}, {Poulin}, {Derome} et~al.}]{2017PhRvL.119x1101G}
{Génolini} Y, {Serpico} PD, {Boudaud} M, {Caroff} S, {Poulin} V, {Derome} L,
  et~al.
\newblock {Indications for a High-Rigidity Break in the Cosmic-Ray Diffusion
  Coefficient}.
\newblock {\em \prl\/} {\bf 119} (2017{\natexlab{a}}) 241101.
\newblock \doi{10.1103/PhysRevLett.119.241101}.

\bibitem[{Vecchi et~al.(2020)}]{Vecchi:2020nvb}
Vecchi M, et~al.
\newblock {Is the B/C slope in AMS-02 data actually telling us something about
  the diffusion coefficient slope?}
\newblock {\em PoS\/} {\bf ICRC2019} (2020) 145.
\newblock \doi{10.22323/1.358.0145}.

\bibitem[{{Maurin}(2020)}]{2020CoPhC.24706942M}
{Maurin} D.
\newblock {USINE: Semi-analytical models for Galactic cosmic-ray propagation}.
\newblock {\em Computer Physics Communications\/} {\bf 247} (2020) 106942.
\newblock \doi{10.1016/j.cpc.2019.106942}.

\bibitem[{Amato and Blasi(2018)}]{Amato:2017dbs}
Amato E, Blasi P.
\newblock {Cosmic ray transport in the Galaxy: A review}.
\newblock {\em Adv. Space Res.\/} {\bf 62} (2018) 2731--2749.
\newblock \doi{10.1016/j.asr.2017.04.019}.

\bibitem[{{Génolini} et~al.(2019){Génolini}, {Boudaud}, {Batista}, {Caroff},
  {Derome}, {Lavalle} et~al.}]{2019PhRvD..99l3028G}
{Génolini} Y, {Boudaud} M, {Batista} PI, {Caroff} S, {Derome} L, {Lavalle} J,
  et~al.
\newblock {Cosmic-ray transport from AMS-02 boron to carbon ratio data:
  Benchmark models and interpretation}.
\newblock {\em \prd\/} {\bf 99} (2019) 123028.
\newblock \doi{10.1103/PhysRevD.99.123028}.

\bibitem[{{Weinrich} et~al.(2020{\natexlab{a}}){Weinrich}, {Boudaud}, {Derome},
  {Génolini}, {Lavalle}, {Maurin} et~al.}]{2020A&A...639A..74W}
{Weinrich} N, {Boudaud} M, {Derome} L, {Génolini} Y, {Lavalle} J, {Maurin} D,
  et~al.
\newblock {Galactic halo size in the light of recent AMS-02 data}.
\newblock {\em \aap\/} {\bf 639} (2020{\natexlab{a}}) A74.
\newblock \doi{10.1051/0004-6361/202038064}.

\bibitem[{{Derome} et~al.(2019){Derome}, {Maurin}, {Salati}, {Boudaud},
  {Génolini}, and {Kunz{\'e}}}]{2019A&A...627A.158D}
{Derome} L, {Maurin} D, {Salati} P, {Boudaud} M, {Génolini} Y, {Kunz{\'e}} P.
\newblock {Fitting B/C cosmic-ray data in the AMS-02 era: a cookbook. Model
  numerical precision, data covariance matrix of errors, cross-section nuisance
  parameters, and mock data}.
\newblock {\em \aap\/} {\bf 627} (2019) A158.
\newblock \doi{10.1051/0004-6361/201935717}.

\bibitem[{Maurin et~al.(2022)Maurin, Bueno, G\'enolini, Derome, and
  Vecchi}]{Maurin:2022irz}
Maurin D, Bueno EF, G\'enolini Y, Derome L, Vecchi M.
\newblock {On the importance of Fe fragmentation for LiBeB analyses: Do we need
  a Li primary source to explain AMS-02 data?}  (2022).

\bibitem[{{Boudaud} et~al.(2020){Boudaud}, {Génolini}, {Derome}, {Lavalle},
  {Maurin}, {Salati} et~al.}]{2020PhRvR...2b3022B}
{Boudaud} M, {Génolini} Y, {Derome} L, {Lavalle} J, {Maurin} D, {Salati} P,
  et~al.
\newblock {AMS-02 antiprotons' consistency with a secondary astrophysical
  origin}.
\newblock {\em Physical Review Research\/} {\bf 2} (2020) 023022.
\newblock \doi{10.1103/PhysRevResearch.2.023022}.

\bibitem[{{Weinrich} et~al.(2020{\natexlab{b}}){Weinrich}, {Génolini},
  {Boudaud}, {Derome}, and {Maurin}}]{2020A&A...639A.131W}
{Weinrich} N, {Génolini} Y, {Boudaud} M, {Derome} L, {Maurin} D.
\newblock {Combined analysis of AMS-02 (Li,Be,B)/C, N/O, $^{3}$He, and $^{4}$He
  data}.
\newblock {\em \aap\/} {\bf 639} (2020{\natexlab{b}}) A131.
\newblock \doi{10.1051/0004-6361/202037875}.

\bibitem[{{Maurin} et~al.(2014){Maurin}, {Melot}, and
  {Taillet}}]{2014A&A...569A..32M}
{Maurin} D, {Melot} F, {Taillet} R.
\newblock {A database of charged cosmic rays}.
\newblock {\em \aap\/} {\bf 569} (2014) A32.
\newblock \doi{10.1051/0004-6361/201321344}.

\bibitem[{Maurin et~al.(2020)Maurin, Dembinski, Gonzalez, Maris, and
  Melot}]{Maurin:2020teo}
Maurin D, Dembinski H, Gonzalez J, Maris IC, Melot F.
\newblock {Cosmic-Ray Database Update: Ultra-High Energy, Ultra-Heavy, and
  Antinuclei Cosmic-Ray Data (CRDB v4.0)}.
\newblock {\em Universe\/} {\bf 6} (2020) 102.
\newblock \doi{10.3390/universe6080102}.

\bibitem[{Ghelfi et~al.(2017)Ghelfi, Maurin, Cheminet, Derome, Hubert, and
  Melot}]{Ghelfi:2016pcv}
Ghelfi A, Maurin D, Cheminet A, Derome L, Hubert G, Melot F.
\newblock {Neutron monitors and muon detectors for solar modulation studies: 2.
  \ensuremath{\phi} time series}.
\newblock {\em Adv. Space Res.\/} {\bf 60} (2017) 833--847.
\newblock \doi{10.1016/j.asr.2016.06.027}.

\bibitem[{{Mertsch}(2020)}]{2020Ap&SS.365..135M}
{Mertsch} P.
\newblock {Test particle simulations of cosmic rays}.
\newblock {\em \apss\/} {\bf 365} (2020) 135.
\newblock \doi{10.1007/s10509-020-03832-3}.

\bibitem[{{Kraichnan}(1965)}]{1965PhFl....8.1385K}
{Kraichnan} RH.
\newblock {Inertial-Range Spectrum of Hydromagnetic Turbulence}.
\newblock {\em Physics of Fluids\/} {\bf 8} (1965) 1385--1387.
\newblock \doi{10.1063/1.1761412}.

\bibitem[{{Kolmogorov}(1941)}]{1941DoSSR..30..301K}
{Kolmogorov} A.
\newblock {The Local Structure of Turbulence in Incompressible Viscous Fluid
  for Very Large Reynolds' Numbers}.
\newblock {\em Akademiia Nauk SSSR Doklady\/} {\bf 30} (1941) 301--305.

\bibitem[{{Génolini} et~al.(2018){Génolini}, {Maurin}, {Moskalenko}, and
  {Unger}}]{2018PhRvC..98c4611G}
{Génolini} Y, {Maurin} D, {Moskalenko} IV, {Unger} M.
\newblock {Current status and desired precision of the isotopic production
  cross sections relevant to astrophysics of cosmic rays: Li, Be, B, C, and N}.
\newblock {\em \prc\/} {\bf 98} (2018) 034611.
\newblock \doi{10.1103/PhysRevC.98.034611}.

\bibitem[{{Schroer} et~al.(2021){Schroer}, {Evoli}, and
  {Blasi}}]{2021PhRvD.103l3010S}
{Schroer} B, {Evoli} C, {Blasi} P.
\newblock {Intermediate-mass and heavy Galactic cosmic-ray nuclei: The case of
  new AMS-02 measurements}.
\newblock {\em \prd\/} {\bf 103} (2021) 123010.
\newblock \doi{10.1103/PhysRevD.103.123010}.

\bibitem[{{Boschini} et~al.(2021){Boschini}, {Della Torre}, {Gervasi},
  {Grandi}, {J{\'o}hannesson}, {La Vacca} et~al.}]{2021ApJ...913....5B}
{Boschini} MJ, {Della Torre} S, {Gervasi} M, {Grandi} D, {J{\'o}hannesson} G,
  {La Vacca} G, et~al.
\newblock {The Discovery of a Low-energy Excess in Cosmic-Ray Iron: Evidence of
  the Past Supernova Activity in the Local Bubble}.
\newblock {\em \apj\/} {\bf 913} (2021) 5.
\newblock \doi{10.3847/1538-4357/abf11c}.

\bibitem[{{Génolini} et~al.(2017{\natexlab{b}}){Génolini}, {Salati},
  {Serpico}, and {Taillet}}]{2017A&A...600A..68G}
{Génolini} Y, {Salati} P, {Serpico} PD, {Taillet} R.
\newblock {Stable laws and cosmic ray physics}.
\newblock {\em \aap\/} {\bf 600} (2017{\natexlab{b}}) A68.
\newblock \doi{10.1051/0004-6361/201629903}.

\bibitem[{{Salati} et~al.(2017){Salati}, {G{\'e}nolini}, {Serpico}, and
  {Taillet}}]{2017EPJWC.13602006S}
{Salati} P, {G{\'e}nolini} Y, {Serpico} P, {Taillet} R.
\newblock {The proton and helium anomalies in the light of the Myriad model}.
\newblock {\em European Physical Journal Web of Conferences\/} (2017), {\em
  European Physical Journal Web of Conferences\/}, vol. 136, 02006.
\newblock \doi{10.1051/epjconf/201713602006}.

\bibitem[{{Porter} et~al.(2019){Porter}, {J{\'o}hannesson}, and
  {Moskalenko}}]{2019ApJ...887..250P}
{Porter} TA, {J{\'o}hannesson} G, {Moskalenko} IV.
\newblock {Deciphering Residual Emissions: Time-dependent Models for the
  Nonthermal Interstellar Radiation from the Milky Way}.
\newblock {\em \apj\/} {\bf 887} (2019) 250.
\newblock \doi{10.3847/1538-4357/ab5961}.

\bibitem[{{Evoli} et~al.(2021){Evoli}, {Amato}, {Blasi}, and
  {Aloisio}}]{2021PhRvD.104l3029E}
{Evoli} C, {Amato} E, {Blasi} P, {Aloisio} R.
\newblock {Stochastic nature of Galactic cosmic-ray sources}.
\newblock {\em \prd\/} {\bf 104} (2021) 123029.
\newblock \doi{10.1103/PhysRevD.104.123029}.

\bibitem[{Chang et~al.(2017)}]{DAMPE:2017cev}
Chang J, et~al.
\newblock {The DArk Matter Particle Explorer mission}.
\newblock {\em Astropart. Phys.\/} {\bf 95} (2017) 6--24.
\newblock \doi{10.1016/j.astropartphys.2017.08.005}.

\bibitem[{Torii and Marrocchesi(2019)}]{Torii:2019wxn}
Torii S, Marrocchesi PS.
\newblock {The CALorimetric Electron Telescope (CALET) on the International
  Space Station}.
\newblock {\em Adv. Space Res.\/} {\bf 64} (2019) 2531--2537.
\newblock \doi{10.1016/j.asr.2019.04.013}.

\bibitem[{Atkin et~al.(2017)}]{Atkin:2017vhi}
Atkin E, et~al.
\newblock {First results of the cosmic ray NUCLEON experiment}.
\newblock {\em JCAP\/} {\bf 07} (2017) 020.
\newblock \doi{10.1088/1475-7516/2017/07/020}.

\end{thebibliography}
\end{document}